# On the distributions of restriction sites in human and pangolin sarbecoviruses

Zach Hensel[1]

## Affiliations

[1] ITQB NOVA, Universidade NOVA de Lisboa, Av. da República, Oeiras, Lisboa 2780-157, Portugal

Contact: zach.hensel@itqb.unl.pt

## Abstract

Since early 2020, several theories have suggested that a distribution of restriction endonuclease recognition sites in the SARS-CoV-2 genome indicates a synthetic origin. The most influential of these, a 2022 preprint by Bruttel *et al.* claimed: “The BsaI/BsmBI restriction map of SARS-CoV-2 is unlike any wild-type coronavirus, and it is unlikely to evolve from its closest relatives.” To test this, I reanalyzed the same 11 contested sites using an expanded set of sarbecovirus genomes, including bat coronaviruses published after the Bruttel *et al.* preprint. For each site, I identified the bat coronaviruses most closely matching SARS-CoV-2 in the surrounding sequences, excluding the sites themselves. The Bruttel *et al.* hypothesis predicts that these closely related viruses should differ from SARS-CoV-2 at many of the contested sites if restriction sites had been artificially introduced or removed. Contrary to this prediction, one or more of the most closely related bat coronaviruses are identical to SARS-CoV-2 at all 11 sites. Equivalent “synthetic fingerprints” were identified in natural pangolin sarbecoviruses. Finally, I conducted a re-analysis of the dataset that Bruttel *et al.* used to test where the SARS-CoV-2 BsaI/BsmBI restriction map was significantly more “evenly spaced” than expected in a natural genome. I found technical and conceptual errors that resulted in Bruttel *et al.* reporting that their chosen metric was 0.07% likely to occur by chance rather than 4.2%, reducing the apparent rarity 60-fold. Using a more informative metric, I tested whether restriction sites in SARS-CoV-2 or two pangolin sarbecoviruses were significantly more evenly spaced than expected and found they were not. These results show that the restriction maps of SARS-CoV-2 and related pangolin viruses are unremarkable in the context of related bat coronaviruses.

## Changelog

- Version 2
    - Added short results section describing misrepresentation of a previous paper by Bruttel *et al.*
    - Added **Supplementary Note 2** responding to post-publication criticism

# Introduction

A 2022 preprint by Bruttel *et al.* claimed to have found a "synthetic fingerprint" in the SARS-CoV-2 genome, suggesting a lab origin (Bruttel, Washburne and VanDongen, 2023). They argued that the pattern of BsaI and BsmBI recognition sites (the SARS-CoV-2 BsaI/BsmBI restriction map) was unnatural and likely engineered.

Instead, Bruttel *et al.* hypothesized that a bat sarbecovirus genome such as RmBANAL52 had been modified by adding and removing BsaI and BsmBI sites in order to produce a construct with what, they argued, was an extremely unusual distribution of restriction sites. The preprint concluded that the BsaI/BsmBI restriction map in SARS-CoV-2 "is unlikely to evolve from its closest relatives by chance." Their hypothesis was illustrated as "How to make SARS2" in their preprint. Bruttel *et al.* identified 11 sites in the SARS-CoV-2 genome that they concluded may have been edited to add or remove restriction sites.

Others immediately raised sound methodological arguments for why this was an implausible hypothesis. Phylogenetic analysis found that "65/66 of the nucleotides in restriction enzyme cut-sites were already shared by the ancestor of SARS-CoV-2 and natural bat viruses (Crits-Christoph and Pekar, 2022)." One study noted that the BsaI/BsmBI restriction maps of some other coronaviruses have similar characteristics (Wu, 2023). Some pre- and post-publication criticism of Bruttel *et al.* is summarized in **Supplementary Note 1**. Notably, Bruttel *et al.* did not consider the mosaic nature of the SARS-CoV-2 genome. This results from extensive recombination between related coronaviruses, with different regions of the genome varying in their ancestry (Boni *et al.*, 2020; Lam *et al.*, 2020).

The specific hypothesis from Bruttel *et al.* continues to be cited in publications without accounting for all the evidence against it. One recent analysis of SARS-CoV-2 origins gave significant weight to results from Bruttel *et al.* and did not note any contradictory evidence (Chen *et al.*, 2024). A recent review did not note the most compelling evidence against the Bruttel *et al.* hypothesis (Domingo, 2024). A widely reported white paper concluded, "The presence of restriction enzymes used in coronavirus-related genetic engineering further supports the likelihood of synthetic manipulation (Kadlec, 2025)." And a recently published book uncritically repeated the conclusion that the SARS-CoV-2 BsaI/BsmBI restriction map was unlikely to occur naturally (Bratlie, 2025).

Thus, it is valuable to examine, in a new way, whether the BsaI/BsmBI restriction map is consistent with natural sarbecovirus evolution.

Bruttel *et al.* wrote that "Our hypothesis that SARS-CoV-2 is a reverse genetics system can be tested." They applied a curious selection of statistical tests (e.g., testing whether the map anomalously lacked a long fragment compared to natural virus, but not considering its anomalously short fragment compared to synthetic viruses). They concluded that the BsaI/BsmBI restriction map was "unlikely to evolve from its closest relatives by chance." Yet, the test of whether SARS-CoV-2 was likely to have evolved from its closest relatives—testing whether SARS-CoV-2 differs unexpectedly from its closest relatives at contested sites—was not done.

The analysis presented here tests the hypothesis of Bruttel *et al.* using updated sarbecovirus genomes and finds no evidence supporting a synthetic origin. I show that some of **the closest relatives to SARS-CoV-2 are, in fact, *identical* to SARS-CoV-2 at all contested sites**. No further change was necessary—not in nature and certainly not in a lab. Further, my re-analysis of the dataset used by Bruttel *et al.* identified technical and conceptual errors underlying their finding of an "under 0.07% chance of observing such an [anomalously evenly spaced] restriction map." In fact, **spacing between restriction sites for SARS-CoV-2 and two pangolin sarbecoviruses is not significantly different from that of other coronaviruses in their dataset**.

# Results

## Updated analysis of contested sites

The synthetic origin hypothesis predicts that at the 11 contested BsaI/BsmBI sites, the 6-nucleotide sequence of SARS-CoV-2 will often differ from that of its most closely related bat viruses in the surrounding regions. I tested this prediction using an updated dataset of animal coronavirus genomes, including many published since the Bruttel *et al.* preprint. The natural origin hypothesis predicts they will often be identical. The results, summarized in **Table 1**, unambiguously contradict the conclusion made by Bruttel *et al.* that "The BsaI/BsmBI restriction map of SARS-CoV-2 … is unlikely to evolve from its closest relatives."

| | Sequences most similar to SARS-CoV-2 in region surrounding contested sites (rank) | | | Number of identical sequences at contested sites | |
|---|---|---|---|---|---|
| Position | 1 | 2 | 3 | Top 3 | Ou *et al.* 2025 |
| 2193-2198 | **RpBANAL103** | **RpYN06** | **RacCS203** | 3 | 2 |
| 9751-9756 | **RpYN06** | **RmYN02** | **RpPrC31** | 3 | 28 |
| 10444-10449 | **RpYN06** | RpBANAL103 | RmYN02 | 1 | 0 |
| 11648-11653 | RaTG13 | RacCS203 | **RpPrC31** | 1 | 0 |
| 17329-17334 | **Ra22QT77** | **RshSTT200** | **RmBANAL247** | 3 | 32 |
| 17972-17977 | **RpYN06** | **RmBANAL247** | RmYN02 | 2 | 0 |
| 22922-22927 | **RmBANAL52** | **RmaBANAL236** | **RpBANAL103** | 3 | 0 |
| 22923-22928 | **RmBANAL52** | **RmaBANAL236** | **RpBANAL103** | 3 | 0 |
| 23292-23297 | **RmBANAL52** | **RmaBANAL236** | **RpBANAL103** | 3 | 6 |
| 24102-24107 | **Rp22DB159** | **RmBtSY2** | RshSTT200 | 2 | 1 |
| 24509-24514 | **RmBANAL52** | **RmaBANAL236** | **RpBANAL103** | 3 | 7 |

**Table 1.** The three genomes with the highest similarity to SARS-CoV-2 in the region around, *but not including*, the 6-nucleotide genome position of interest are shown ranked from highest to lowest identity to SARS-CoV-2. The names of genome sequences that are identical to SARS-CoV-2 at all six nucleotides are colored green. The last two columns show the number of times the site sequence is identical to SARS-CoV-2 in the three most similar sequences and also in 33 recently reported bat coronavirus genomes from Cambodia (Ou et al., 2025).

Of the 11 sites analyzed, **all 11 have a perfect, 6-nucleotide match** in at least one of the top three most closely related bat coronaviruses (based on similarity of nucleotides surrounding, but not including, the 6-nucleotide sites). The regions analyzed started from the 100 nucleotides around each site (50 on each side) and were expanded symmetrically until there were only three related viruses with the highest local similarity. **Figure 1** shows sequence alignments for one site present in SARS-CoV-2 and absent in bat coronaviruses RmBANAL52 and RaTG13. There are no sites for which the SARS-CoV-2 sequence is anomalous. See **Data Supplement 1** for full results.

```
Position 24102-24107: SARSCoV2 (present), RmBANAL247 (present), RmBANAL52 (absent), RaTG13 (absent)
Top match #1: Rp22DB159 (98.0% identity in 100 nucleotides around site)
SARSCoV2       : CCTTGGTGATATTGCTGCTAGAGACCTCATTTGTGCACAAAAGTTT
                 **********************************************
Rp22DB159      : CCTTGGTGATATTGCTGCTAGAGACCTCATTTGTGCACAAAAGTTT

Top match #2: RmBtSY2 (98.0% identity in 100 nucleotides around site)
SARSCoV2       : CCTTGGTGATATTGCTGCTAGAGACCTCATTTGTGCACAAAAGTTT
                 **********************************************
RmBtSY2        : CCTTGGTGATATTGCTGCTAGAGACCTCATTTGTGCACAAAAGTTT

Top match #3: RshSTT200 (97.0% identity in 100 nucleotides around site)
SARSCoV2       : CCTTGGTGATATTGCTGCTAGAGACCTCATTTGTGCACAAAAGTTT
                 ************************ *********************
RshSTT200      : CCTTGGTGATATTGCTGCTAGAGATCTCATTTGTGCACAAAAGTTT
```

**Figure 1.** Results for one contested BsaI site in SARS-CoV-2 (blue). An identical sequence is found in two closely related bat coronavirus sequences, Rp22DB159 (Vietnam) and RmBtSY2 (China) that each have 98% identity to SARS-CoV-2 in the 100 nucleotides surrounding, but not including, the 6-nucleotide site. Only 20 nucleotides on each side are shown here.

The distribution of these identical sequences among the top three relatives is as follows:

- For seven of the 11 sites, all three of the viruses that have the highest identity to SARS-CoV-2 in the sequence around, but not including, the 6-nucleotide restriction site share the exact same 6-nucleotide sequence as SARS-CoV-2.
- For two of the 11 sites, two of the three closest relatives are identical to SARS-CoV-2 at the site.
- For two of the 11 sites, one of the three closest relatives is identical to SARS-CoV-2.
- None of the 11 sites lacked an identical match.

## Significance of newly published bat coronavirus genomes

This result is unsurprising in light of the phylogenetic analysis discussed in the introduction, which found that the reconstructed ancestral sequence of SARS-CoV-2 and bat coronaviruses was identical to SARS-CoV-2 at 65 of the 66 positions in the 11 contested sites (Crits-Christoph and Pekar, 2022). However, that analysis, conducted in 2022, failed to find a closely related virus with the same BsaI site at position 24,102–24,107 as SARS-CoV-2. **Table 1** shows that this BsaI site is now found in RmBtSY2 and Rp22DB159, which were published later (Wang *et al.*, 2023; Hassanin *et al.*, 2024). A phylogenetic tree of the non-recombinant region including position 24,102–24,107 is reproduced below (**Figure 2**), showing that these two viruses are closely related to SARS-CoV-2 in this region. The BsaI/BsmBI restriction map of SARS-CoV-2 is also identical to that of the inferred recombinant common ancestor, RecCA, in the most recent analysis including RmBtSY2 and Rp22DB159 (Pekar *et al.*, 2025).

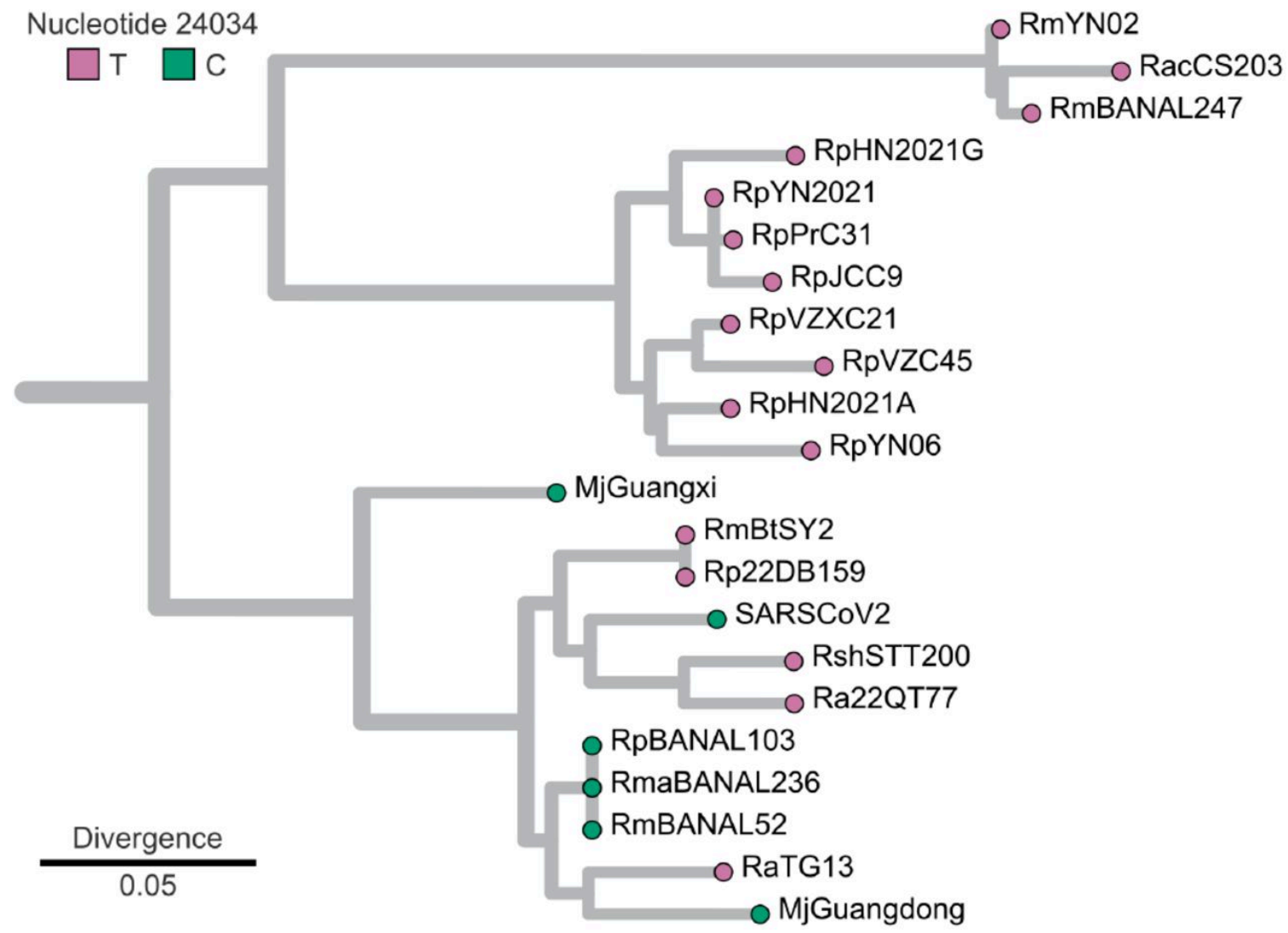

**Figure 2.** Reproduction of Figure 5 from a recent preprint (Hensel and Débarre, 2025), which was made available under the CC-BY-NC-ND 4.0 International license. A phylogenetic tree is depicted for the non-recombinant region corresponding to positions 23734–24183 in the SARS-CoV-2 reference genome. Recently published sequences for RmBtSY2 and Rp22DB159 are the most identical to SARS-CoV-2 in the region surrounding the contested site 24102–24107, and are also identical to SARS-CoV-2 within this region. Here, sequences are colored by the identity at position 24034, demonstrating uncertainty in identifying recombination breakpoints.

Significant new data continue to accumulate, unsurprisingly confirming the natural origin of the sequence at the 11 contested sites. The final column in **Table 1** shows how often the 33 bat coronavirus genomes from Cambodia reported in a recent preprint (Ou *et al.*, 2025) are identical to SARS-CoV-2 at these sites. The BsaI site at position 24102–24107 is found in one, RmaSTT500, which is closely related to SARS-CoV-2 in this region. See **Data Supplement 2** for the complete list of new Cambodian sequences identical at each contested site.

## Evaluation of the restriction site hypothesis for pangolin coronaviruses

The methodology of Bruttel *et al.* is persuasive only if one assumes its specificity—that equivalent criteria are unlikely to exist that can find a "synthetic fingerprint" in a natural virus. To test this assumption, I chose the genome MP789 (Liu *et al.*, 2020) as a test case, a SARS-CoV-2-like coronavirus from a pangolin. I will refer to this sample as MjGuangdong rather than MP789 in this manuscript following a consistent naming convention (Hassanin *et al.*, 2024) reflecting species and sample location (Mj: *Manis javanica*; Guangdong province). The earliest report of viral sequences from this sample from smuggled pangolins predates the COVID-19 pandemic (Liu, Chen and Chen, 2019) and there is no doubt that it is a natural virus. Like SARS-CoV-2, this virus is phylogenetically distinct, lacking a relative in bats with very high identity across its entire genome.

I examined the BsaI and BsmBI restriction map for MjGuangdong, and I found a "synthetic fingerprint" in MjGuangdong that meets the criteria by Bruttel *et al.* for SARS-CoV-2 (5–7 fragments upon digest with a type IIS restriction enzyme, each less than 8 kb, unique overhangs). Digest with BsaI produces 7 fragments ranging from 495 to 7,552 nucleotides. In fact, although MjGuangdong is a natural genome, the same analysis that I applied above for SARS-CoV-2 fails to identify some sites in the false "synthetic fingerprint" in its most closely related bat coronavirus genomes. Three BsaI sites in MjGuangdong are missing in closely related viruses, and one BsaI missing from MjGuangdong is found in the most closely related coronaviruses. See **Data Supplement 3** for full results.

So, does this mean we should reconsider whether MjGuangdong is artificial? Of course not. Examination of the pairwise identity between MjGuangdong and its most closely related viruses (average highest identity 92.5% in the region around the 10 sites investigated; 93.6% for 6 sites with one or more identical matches and 90.8% for 4 sites without identical matches) and SARS-CoV-2 (98.1%) explains the difference. MjGuangdong is simply more different from other viruses published to date than is the case for SARS-CoV-2, so identical sequences at these sites are less likely to have been found in other published viral genomes. Further, as with SARS-CoV-2, the observed restriction map was not used when recovering MjGuangdong using reverse genetics, showing that these are not convenient reverse genetics systems at all (Hou *et al.*, 2023).

Instead, this exercise is a cautionary tale about the perils of seeing patterns in high-dimensional genomic data. It only demonstrates how any sufficiently complex dataset can be made to yield a seemingly significant, yet ultimately meaningless, result. For a final test, I considered the other pangolin sarbecovirus with a high-quality, full-length genome available, MjGuangxi (Lam *et al.*, 2020). Again, I identified a "synthetic fingerprint" meeting the Bruttel *et al.* requirements. Digestion with PaqCI produces 7 fragments ranging from 1,515 to 6,843 nucleotides. MjGuangxi is even less likely than MjGuangdong to match its most closely related sequences at sites of interest, matching at only 3 of 10 sites (**Data Supplement 4**). Again, this is a natural virus and this is not a sign of engineering; it reflects a combination of the degree of divergence between MjGuangxi and the most closely related viruses that have been sequenced, as well as PaqCI having a 7-nucleotide rather than a 6-nucleotide recognition sequence that is more likely to include a mismatch.

## Re-analysis of restriction site spacing

Bruttel *et al.* not only concluded that the SARS-CoV-2 BsaI/BsmBI restriction map was anomalously different from RmBANAL52 and other closely related bat coronaviruses. They also argued that "evenly spaced restriction sites" were a hallmark of synthetic coronaviruses in general, finding that the longest fragment of the SARS-CoV-2 BsaI/BsmBI restriction map (7,578 nucleotides) was the most anomalously short out of 72 coronavirus genomes subjected to digestion by six type IIS restriction enzymes and their unique combinations. I conducted a re-analysis of their dataset (72 coronaviruses) using their methodology, which included BglI (not a type IIS enzyme), and reproduced the result that the longest fragment of the SARS-CoV-2 BsaI/BsmBI restriction map was the most anomalously short of all longest fragments for the subset of restriction maps with 5–7 fragments. I corrected two methodological errors: (1) one coding error that greatly inflated the denominator when estimating significance (see below) and (2) double-counting single digests in cases where one enzyme in a pair cut zero times. I also included PaqCI in my primary analysis, which was not considered by Bruttel *et al.* for technical reasons. This resulted in a total of 2,016 possible combinations of enzymes (single- or double-digest) and viral genomes, with 1,915 digestions calculated owing to the absence of recognition sites for some enzymes in some genomes.

The 72-genome dataset used by Bruttel *et al.* also includes multiple subsets of identical or nearly identical genomes (e.g., two copies of SARS-CoV Tor2 plus one copy of 99.98% identical SARS-CoV Urbani, two 99.94% identical TGEV sequences, etc). However, I retained most duplicates and near duplicates rather than rigorously deduplicating the dataset. The only exception, following Bruttel *et al.*, was removing restriction maps for SARS-CoV-2 sequence WIV04 (GenBank MN996528), which is identical to the reference sequence Hu-1 (GenBank NC_045512.2). Exclusion of this sequence reduced the dataset to 71 sequences with 1,887 unique restriction maps (inclusive of 17 single-enzyme maps with no cuts), 308 with 5–7 fragments.

Since Bruttel *et al.* conducted post hoc analysis of the SARS-CoV-2 BsaI/BsmBI restriction map *after* observing that it fit criteria that they associated with synthetic viruses, it was not appropriate to repeat the analysis described in their preprint: comparing a *single* SARS-CoV-2 restriction map to *every* other restriction map with 5–7 fragments. Instead, for each viral genome I chose the restriction map *best* matching their criteria: 5–7 evenly spaced fragments. To select for even spacing, I chose the restriction map with the smallest coefficient of variation in fragment size. All 71 viral genomes have at least one restriction map with 5–7 fragments. 16 of 71 (23%) of the genomes have a restriction map meeting all of Bruttel *et al.*'s criteria (5–7 fragments; longest fragment shorter than 8 kb). This increases to 28 of 71 (39%) if considering 8-fragment digests, which in fact was the criterion listed by Bruttel *et al.*, but not used for this analysis. Thus, SARS-CoV-2 is unremarkable in meeting these criteria. The most evenly spaced restriction maps for SARS-CoV-2, MjGuangdong, and MjGuangxi were the same ones manually identified above. **Figure 3** shows how SARS-CoV-2 and the pangolin sarbecoviruses compared to the other 68 genomes in terms of being "evenly spaced."

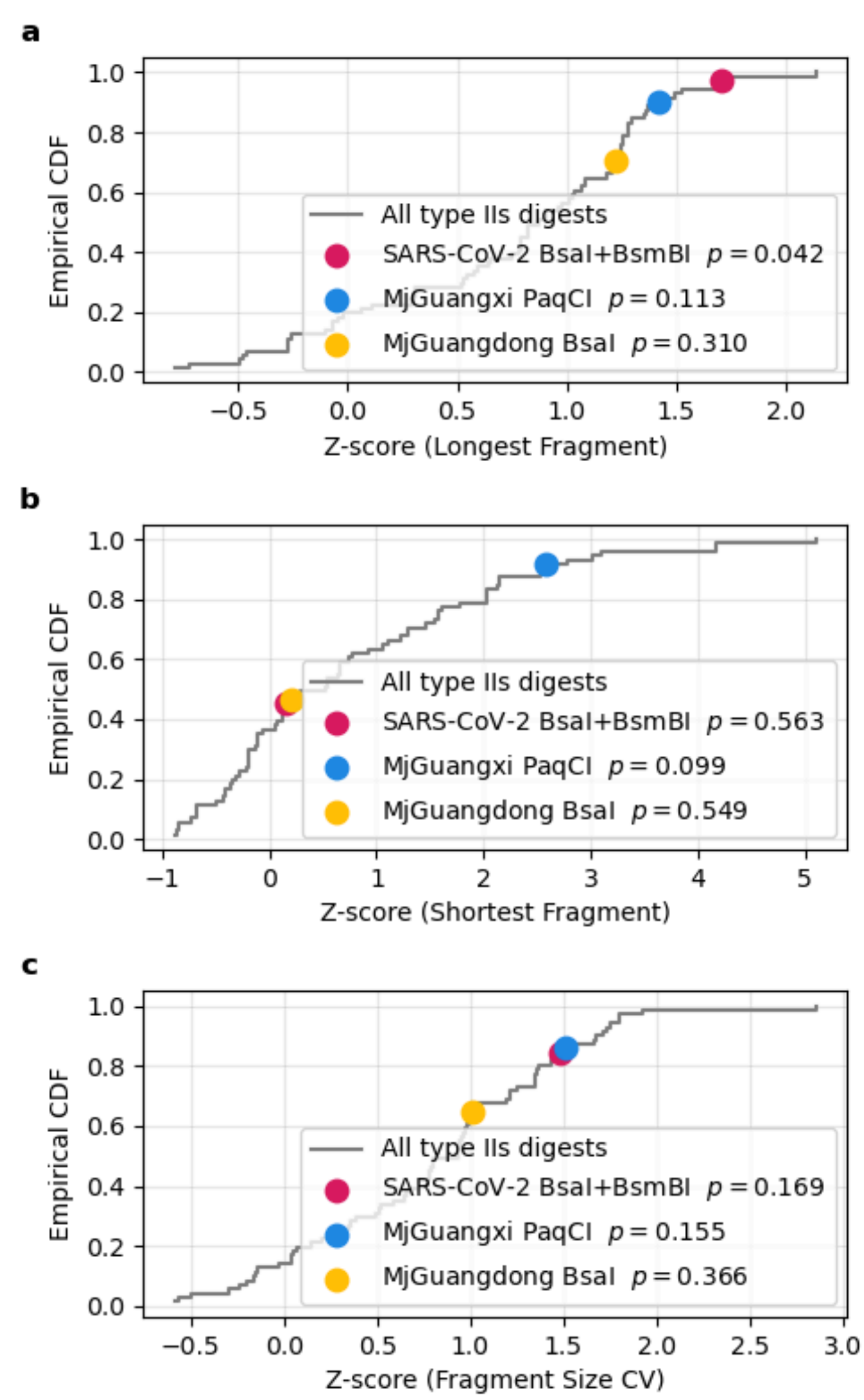

**Figure 3.** Comparison of different metrics quantifying the degree of "evenly spaced" sites for SARS-CoV-2 and two pangolin sarbecoviruses in the context of restriction maps of 71 coronavirus genomes. For each genome, the restriction map from a single- or double-digest with type IIS enzymes which gives 5–7 fragments and the lowest coefficient of variation of fragment lengths is shown. In all cases, a higher normalized Z-score indicates a more "evenly spaced" restriction map. P-values indicate the fraction of genomes with at least as high a Z-score. Z-scores quantify how each genome's restriction map deviates from the expected mean for (**a**) longest fragment length, (**b**) shortest fragment length, and (**c**) coefficient of variation of fragment lengths.

**Figure 3A** qualitatively reproduces the result from Bruttel *et al.*, suggesting that the longest fragment of the SARS-CoV-2 BsaI/BsmBI map is an anomaly. However, now 4.2% of comparable maps have longest fragments as long as or longer than that in SARS-CoV-2 (compared by a normalized Z-score; see **Methods**). This is much less of an apparent anomaly than the 0.07% reported by Bruttel *et al.* The 0.07% calculation by Bruttel *et al.* arose primarily from a coding error and should have been 0.45% (the first-ranked position of SARS-CoV-2 was divided by 1,491, the total number of restriction maps, rather than 221, the number with 5–7 fragments, as described in the paper). The remaining difference between 4.2% and 0.45% owes to three factors:

1. Occasionally double-counting single-digests in cases where one of two enzymes in a double-digest does not cut, as described above. Correcting this technical error changes the likelihood from 0.45% to 0.50%.
2. The SARS-CoV-2 BsaI/BsmBI restriction map is not the only one resulting in 5–7 fragments. Digestion with BsmBI and BglI also produces 5 fragments. Retaining only the restriction map with the highest longest-fragment Z-score for *all* genomes is a necessary balance to compare SARS-CoV-2 to 70 other genomes on equal terms. Correcting this conceptual error changes the likelihood from 0.50% to 1.45%.
3. The Bruttel *et al.* analysis code specified inclusion of PaqCI, but this was not supported by an R library used in the analysis. Since the analysis was premised on type IIS enzymes used in reverse genetics systems, PaqCI should have been included in the analysis since, for example, its isoschizomer AarI was used to construct icSARS-CoV (Yount *et al.*, 2003). In my re-analysis without this technical limitation, the likelihood increases from 1.45% to 4.23%.

However, more importantly, the length of a single fragment is a poor measure of how evenly spaced a restriction map is. **Figure 3B** shows that the shortest fragment in MjGuangxi is somewhat longer than expected from the distribution in other natural viruses, which would also be an expectation of a piecewise synthetic virus (see note 2 in **Supplementary Note 1**). A more appropriate comparison is shown in **Figure 3C**, which shows that the variation in fragment lengths for all three of SARS-CoV-2, MjGuangdong, and MjGuangxi is not significantly anomalous. There is no sign that restriction maps for SARS-CoV-2 or these two pangolin sarbecoviruses are significantly more "evenly spaced" than expected.

## Misinterpretation of previous work

One criteria of the "IVGA fingerprint" described by Bruttel *et al.* is "Two unique recognition sites may flank regions meant to be further manipulated." I did not consider this in the analysis above, because its inclusion in the "fingerprint" arises from misinterpretation of work described in a previous paper. Bruttel *et al.* write that this work "used two distinct endonucleases for genome assembly, with two sites of one enzyme [BsaI] flanking a region of interest, enabling efficient manipulations of the flanking region without having to reassemble the entire viral backbone for each variant." In fact, BsaI sites used for this assembly were *not* retained in the final construct (Hu *et al.*, 2017). Furthermore, infectious clones produced in that work *did* have BsaI sites in the backbone and in other fragments that would make this procedure impossible without additional mutagenesis. In general, I could not find any precedent in the literature on similar reverse genetics constructs for introducing type IIS restriction sites. Bruttel *et al.* exclusively list examples of introducing BglI sites (not type IIS). There are some examples of removing sites to enable one-pot assembly, but the fact that RecCA shares the SARS-CoV-2 BsaI/BsmBI restriction map strongly suggests that there is no history of this for SARS-CoV-2 (Pekar *et al.*, 2025).

# Discussion

Of the 11 sites Bruttel *et al.* proposed to be engineered by adding or removing restriction sites, all 11 have a perfect 6-nucleotide match in at least one of their top three most identical relatives in the regions around, *but not including*, the 6-nucleotide sites. This result is inconsistent with a synthetic origin hypothesis that requires modifying these sites. It is, however, consistent with the hypothesis that a recent ancestor of SARS-CoV-2 circulating in bats was already identical to SARS-CoV-2 at all 11 contested sites.

In the revised version of their preprint, Bruttel *et al.* acknowledged the omission of recombination that was implicit in their recipe for "How to make SARS2" from RmBANAL52: "we considered point mutations and not recombination events." There is no way to accurately discuss how SARS-CoV-2 differs from a virus as different as RmBANAL52 without considering recombination (Temmam *et al.*, 2022). Sarbecovirus genomes are mosaics, reflecting frequent recombination. It is critical to consider recombination when investigating SARS-CoV-2 origins.

What emerges when analyzing SARS-CoV-2 and its relatives in light of recombination is not an enigma, but a natural history that is becoming remarkably clear. Recent analysis estimated that common ancestors of SARS-CoV-2 and closely related bat coronaviruses circulated only six years before 2020 (Pekar et al., 2025) and future work will shed light on what happened between then and the emergence of SARS-CoV-2 in humans in Wuhan, China.

Despite extensive contradictory evidence, the "synthetic fingerprint" hypothesis continues to be cited, justifying this new, simplified approach for comparative genomic analysis. Further, my complete re-analysis of fragment length distributions in the same dataset used by Bruttel *et al.* found that the spacing of BsaI/BsmBI sites in SARS-CoV-2 is not anomalous. In over five years, no laboratory has reported using this so-called ideal system for SARS-CoV-2 research. The fact that I found equivalent "synthetic fingerprints" for both MjGuangdong and MjGuangxi illustrates the possible pitfalls of pattern-finding in genomic data. It suggests that a more patient and clever analyst could construct a genomic engineering hypothesis far more difficult to falsify.

As shown in the **Supplementary Note 1**, Bruttel *et al.* reject critiques of this nature because, they say, data from China might be fabricated. This reasoning reflects selective data exclusion rather than methodological rigor, and it is directly contradicted by the evidence presented here. The crucial bat viruses that are identical to SARS-CoV-2 at contested sites (**Table 1**) were sampled not only in China, but also in Laos, Cambodia, Vietnam, and Thailand, in collaboration with international teams of scientists from other countries.

# Methods

## Sequences and alignments

Analysis was performed on a custom multiple sequence alignment (MSA) of sarbecovirus genomes. The starting point was a previously published MSA (Hassanin *et al.*, 2024) of 111 coronavirus genomes. To this, the genomes of bat coronavirus RmBtSY2 ((Wang *et al.*, 2023) GenBank: OP963576.1) and 33 Cambodian samples ((Ou *et al.*, 2025) GISAID: EPI_ISL_19825786–EPI_ISL_19825818) using the MAFFT server (Katoh and Standley, 2013).

Viral sequences from Cambodia were not used for the primary analysis in **Table 1** to ensure all included genomes were subject to the same filtering criteria as the baseline alignment (Hassanin *et al.*, 2024).

The dataset of 72 viral genomes was obtained from the GitHub repository associated with the manuscript from Bruttel *et al.* at https://github.com/reptalex/SARS2_Reverse_Genetics.

## Analysis of contested restriction sites

A custom Python script was developed to test the hypothesis from Bruttel *et al.* First, a coordinate map was established by performing a global alignment between the SARS-CoV-2 sequence from the MSA and the reference genome (GenBank: NC_045512) to translate all MSA positions into standard reference coordinates.

The script then identified all 11 BsaI/BsmBI sites that were present in SARS-CoV-2 but absent in three closely related viral genomes (RmBANAL247, RmBANAL52, RaTG13), or vice-versa. This was the same criteria used by Bruttel *et al.* in their preprint. This identified the same 11 sites as a previous analysis (Crits-Christoph and Pekar, 2022). For each of these 11 contested, 6-nucleotide sites, the script identified the three genomes in the full MSA most identical to SARS-CoV-2. This comparison was performed on the 100 nucleotides immediately surrounding the site (50 nucleotides upstream and 50 downstream), while the 6-nucleotide site itself was masked and excluded from the identity calculation.

In the event of a tie for the third-ranked genome, the window was expanded symmetrically by one nucleotide on each side and the identities were recalculated. This process was repeated until a unique set of top three most identical genomes was unambiguously identified.

The analysis for MjGuangdong was identical, except that only BsaI was investigated and a different reference sequence was used (GenBank: MT121216). BsaI was chosen based on a preliminary look at the BsaI/BsmBI restriction map and observing that the BsaI restriction map met the "synthetic fingerprint" requirements of Bruttel *et al.* For MjGuangxi analysis, the P2V sequence was used as a reference (GenBank: MT072864). PaqCI was identified as meeting the "synthetic fingerprint" criteria after first observing that a combination of BsaI, BsmBI, and double-digested fragments would also meet the same criteria (an internal BsmBI site in one BsaI-digested fragment would not be problematic in hypothetical reverse genetics system). However, I chose to focus on the single PaqCI restriction map for its simplicity.

## Analysis of restriction map spacing

I based my analysis on the "type IIs" analysis in the script `2_CoV_restriction_mapping.R` in the GitHub repository associated with Bruttel *et al.* available at https://github.com/reptalex/SARS2_Reverse_Genetics. Methodological errors described in **Results** were addressed in a custom Python script. The Z-scores plotted in **Figure 3** depict the deviation from the mean value for a restriction map with a given number of fragments (5, 6, or 7), normalized by the standard deviation of values for the same set of restriction maps. The sign of the Z-score is inverted for **Figure 3B** relative to the other two figures, so that a high Z-score denotes a relatively *long* shortest fragment, while a high Z-score denotes a relatively *short* longest fragment in **Figure 3A** and a relatively *low* coefficient of variation in **Figure 3C**. Following Bruttel *et al.*'s methodology, Z-scores were normalized separately for 5–, 6–, and 7–fragment restriction maps ($n$ = 102, 103, and 103 respectively), and I neglected to consider that genome lengths varied between 25,425 and 31,491 nucleotides, which will of course impact expected fragment lengths. Results excluding PaqCI as a possible restriction enzyme were reproducible in my Python script and using an R script adapted from `2_CoV_restriction_mapping.R`, making changes to correct errors and be consistent with the text of Bruttel *et al.*

(i.e., analysis relative to other 5–7 fragment restriction maps and not double counting some single-digest restriction maps).

Here, I identify specific elements of `2_CoV_restriction_mapping.R` that I addressed.

1. SapI and PaqCI are not included in the R package used by Bruttel *et al.* I did not use SapI (isoschizomer of BspQI). I also note that BtgZI could have been included, but I also excluded it from this analysis. The enzymes that do not contribute to the analysis are shown in red.

```
# Load restriction enzyme set
data("RESTRICTION_ENZYMES")
Type2Res <- c('BbsI','BfuAI','BspQI','PaqCI','SapI','BsaI','BsmBI','BglI')
Type2Res <- Type2Res[(Type2Res %in% names(RESTRICTION_ENZYMES))]
```

2. The line shown in red here, missing in the original code, is required to give the correct denominator for the number of restriction maps with 5–7 fragments.

```
frags_2s <- frags_2s[,list(max_fragment_length=max(fragment_lengths)/genome_length[1],
                           no_fragments=.N,
                           genome_length=genome_length[1],

…

setkey(frags_2s,no_fragments)
setkey(rgs_specs,no_fragments)
frags_2s <- rgs_specs[frags_2s]

…

frags_2s <- frags_2s[no_fragments >= 5 & no_fragments <= 7]

frags_2s[,.N]
## 1491 total # 221 after correctly filtering for 5-7 fragments
```

## Data Availability

No sequence data was generated for this study. All sources of sequence data are described in the Methods section. Alignments cannot be published due to GISAID data sharing restrictions. Outputs of analysis scripts used to populate **Table 1** are reproduced in **Data Supplements 1–4**.

## Acknowledgments

I gratefully acknowledge all data contributors, i.e., the Authors and their Originating laboratories responsible for obtaining the specimens, and their Submitting laboratories for generating the genetic sequence and metadata and sharing via the GISAID Initiative and other public databases, on which this research is based. Z.H. was supported by FCT - Fundação para a Ciência e a Tecnologia, I.P., through MOSTMICRO-ITQB R&D Unit (https://doi.org/10.54499/UIDB/04612/2020; https://doi.org/10.54499/UIDP/04612/2020) and LS4FUTURE Associated Laboratory (https://doi.org/10.54499/LA/P/0087/2020).

# Supplemental Information

## Supplementary Note 1: Pre- and post-publication criticism of Bruttel *et al.*

The 2022 preprint from Bruttel *et al.* was immediately criticized for its hypothesis that employed BsaI/BsmBI in an unconventional way[1] with one inexplicably short, 643-nucleotide fragment[2]. Although some scientists initially found the hypothesis compelling[3], one who conducted further analysis concluded that the BsaI/BsmBI restriction map "can be explained by extensive genetic recombination[4]."

Compelling support for the alternative hypothesis that the 11 contested sites in SARS-CoV-2 had a natural origin was raised in discussions with Bruttel *et al.* a month and a half *before* publishing their preprint, with one critic noting, "The two nucleotides in Wuhan-Hu-1 (10438, 11654) that you have alleged to be perfectly altered to remove Bsal sites exist naturally in RpYN06.[5]" The existence of contested sequences in closely related bat coronaviruses including RpYN06 was also noted in an article published a few weeks after Bruttel *et al.* (Garry, 2022).

However, RpYN06 was not discussed in the original preprint by Bruttel *et al.* nor in a revised version published six months later. In between revisions, Bruttel wrote that "RpYN06 should have been discussed by us[6]" before writing, three days later, "There is just tons of evidence that RaTG13, RpYN06, RmYN02... have been manipulated[7]." What was this evidence? Bruttel initially proposed that RpYN06 was fabricated by Chinese scientists who had seen Bruttel's comments on Twitter[8]. After learning that sequences and sequencing data had been submitted to international databases months *before* his comments, Bruttel wrote, "then the suspicious restriction site pattern. along come [sic] RPyN06 [sic] with some identical sites. none of these viruses were ever sequenced outside of china[9]." Bruttel inexplicably illustrated his reasoning for selectively excluding data from China with a photo of Chinese scientists Yigang Tong and Wuchan Cao, neither of whom are co-authors of the paper reporting RpYN06 (Liu *et al.*, 2020).

In their revised manuscript, Bruttel *et al.* did not describe excluding this contradictory data, nor did they describe any rationale for data exclusion.

---

[1] ***The Economist***. "Scientists dispute a suggestion that SARS-CoV-2 was engineered," 27 October 2022, https://www.economist.com/science-and-technology/2022/10/27/scientists-dispute-a-suggestion-that-sars-cov-2-was-engineered
"Jesse Bloom, an evolutionary virologist at the Fred Hutchinson Cancer Centre in Seattle, says BsmBI is usually used in a way in which the restriction sites involved are eventually removed from the virus."

[2] **Ibid**. "However, Dr. Marillonnet also says that there are arguments against the idea. One is the tiny length of one of the six fragments, something that 'does not seem logical'."

[3] **Ibid**, quoting Dr. Francois Balloux. "Contrary to many of my colleagues, I couldn't identify any fatal flaw in the reasoning and methodology. The distribution of BsaI/BsmBI restriction sites in SARS-CoV-2 is atypical."

[4] **Dr. Francois Balloux** (@BallouxFrancois), Twitter post, 31 October 2022, https://twitter.com/BallouxFrancois/status/1587090507288776704

[5] **Zhihua Chen** (@zhihuachen), Twitter post, 6 September 2022, https://twitter.com/zhihuachen/status/1567233305317199875

[6] **Dr. Valentin Bruttel** (@VBruttel), Twitter post, 7 November 2022, https://twitter.com/VBruttel/status/1589669922942377984

[7] **Dr. Valentin Bruttel (@VBruttel)**, Twitter post, 10 November 2022, https://twitter.com/VBruttel/status/1590730504734912515

[8] **Dr. Valentin Bruttel (@VBruttel)**, Twitter post, 25 October 2022, https://twitter.com/VBruttel/status/1584920770605621248

[9] **Dr. Valentin Bruttel (@VBruttel)**, Twitter post, 5 September 2023, https://twitter.com/VBruttel/status/1698932448141136075

## Supplementary Note 2: Response to post-publication criticism

Co-authors of Bruttel *et al.* have used the website "X" as a platform to respond to this preprint. I will reproduce their posts and respond to them here.

### Valentin Bruttel (@VBruttel)

Dr. Bruttel has a number of comments; I will interleave my responses in red text.

> TL;DR: if you ignore:
>
> a) our main result on elevated mutation rates at Type IIS sites 👇,
> First, Bruttel *et al.* analyzed this *after* noting that BsaI and BsmBI sites differ from high identity genomes e.g., RmBANAL52 and RaTG13. Of course, mutation rates will be higher at sites one has already observed to be different! Second, the relevant comparison is not to genomes with the highest identity across the entire genome but to an ancestral bat virus genome. While, of course, we can't observe the ancestral genome, the analysis I present here is a simplified (and error prone) way to infer its identity. Analysis used to infer RecCA is a more rigorous way to do the same thing (Pekar *et al.*, 2025). Both analyses show that there are not "elevated mutation rates" at these sites, because it's more likely than not that there were no mutations at these sites since the time that a common ancestor existed in bats.
>
> b) DEFUSE plans to build consensus genomes 👇,
> First, this is an argument that does not exist in Bruttel *et al.* so it is not something I addressed in my manuscript. Second, Dr. Bruttel refers to an unfunded research proposal (DEFUSE) and, in doing so, misrepresents the work that was proposed. The proposal reads, "Consensus candidates [sic] genomes will be synthesized…" and cites papers where "consensus" refers to the consensus of sequences of *nearly identical* isolates of the same virus. The purpose is to avoid synthesizing sequencing errors; not to construct some sort of average genome of divergent viruses. Third, hypothetical work constructing such a "consensus" genome (e.g., Bruttel has proposed a "consensus" of RmBANAL247, RmBANAL52, RaTG13, and additional viruses) would be an unprecedented *loss-of-function* experiment discarding decades of evolution. Fourth, this type of "consensus" genome, if it defied the odds and could replicate anywhere, would likely appear obviously synthetic when analyzed since it was not produced by the same mechanisms as natural genomes. Fifth, such a "consensus" genome would not predict the BsaI/BsmBI restriction map in the way that RecCA does (Pekar *et al.*, 2025).
>
> c) the low odds that viruses thousands of miles apart would recombine into a map ideal for modular assembly 👇,
> This argument does not appear in Bruttel *et al.* and the SARS-CoV-2 BsaI/BsmBI map is self-evidently *not* "ideal for molecular assembly" since, to my knowledge, no one has ever used it.
>
> d) references in our preprint where very short fragments are used due to plasmid instability 👇,
> Bruttel is not clear here, so I assume this is meant as a response to explain why the shortest fragment in his proposed SARS-CoV-2 assembly is so short (643 nucleotides). Again, this is not discussed in Bruttel *et al.*, which does not discuss the lengths of shortest fragments at all. Here, Bruttel provides a screenshot from a paper in which one fragment was split into two to address plasmid instability (Zeng *et al.*, 2016). Here, fragment "C" (4,047 nucleotides) was split into fragments "C1" (2,529 nucleotides) and "C2" (1,518 nucleotides). So, the shortest fragment is 136% longer than the shortest fragment in the SARS-CoV-2 BsaI/BsmBI restriction map.
>
> e) documented post-pandemic misrepresentation around SARS-CoV-2–related genomes: inference-review.com/article…
> Here, Bruttel links to an article that suggests, for example, that RaTG13 "might not be the original bat virus but could instead be the result of significant serial passaging of a bat virus in human cells or in mice." First, this is an absurd and baseless accusation of research misconduct. Second, Bruttel *et al.* include RaTG13 in their "broad range of *natural* coronavirus genomes." As described in **Supplementary Note 1** and the **Discussion**, Dr. Bruttel has regularly responded to post-publication criticism by claiming that contradictory data could be fake. Even if there were a rational basis for him to say this (and, let's be clear, there is not), data omission is not described in Bruttel *et al.*
>
> you can arrive at "the restriction site pattern is unusual, but not extremely unusual"
> I show that 39% of the 71 genomes considered by Bruttel *et al.* have the "IVGA fingerprint" of 5–8 fragments in a type IIS restriction map with no fragment longer than 8,000 nucleotides. It is not unusual at all.

### Antonius VanDongen (@tony_vandongen)

Dr. VanDongen notes that, if SARS-CoV-2 inherited its BsaI site from RmBANAL247 via recombination, it would also inherit sites nearby:

> Recombination is not surgical, it tends to swap large fragments (kilobases). That produces a problem for Zach: you inherit spurious mutations that then need to be removed.

This concerns the BsaI site at position 24102-24107 in SARS-CoV-2. Although RmBANAL247 shares this site, it is not one of the genomes most closely related to SARS-CoV-2 in this region (**Figure 2**). As noted in the text of the paper, this site was the *only* contested site not identified in the inferred ancestral genome in a 2022 analysis (Crits-Christoph and Pekar, 2022), but it was subsequently found in closely related viruses Rp22DB159 and BtSY2 and the site is now found in RecCA (Pekar *et al.*, 2025). Neither Rp22DB159 nor BtSY2 have additional BsaI sites near position 24102–24017.

### Michael Weissman (@mbw61567742)

Dr. Weissman, who is not a co-author but who has published a Bayesian analysis in which the "synthetic fingerprint" is one of the strongest pieces of evidence for a synthetic origin, has also responded on X.

His initial review:

> Yeah, read it. Try searching for any form of the word "probability".

Fair enough. My rationale for the analyses that I included is discussed in the **Introduction**. Namely, a rigorous analysis that falsified Bruttel *et al.* in October 2022 (Crits-Christoph and Pekar, 2022) has been ignored in consequential ways. This called for a more accessible approach to the same question.

Asked, "Does the hypothetical RecA [sic] have that spacing?"

> No. The recCA he wants is off by 1 site. Adding the needed ~400 syn muts gives it a 2.8% chance of getting the pattern. Allowing for the likelihood that wasn't the recCA lowers it further, as would allowing for the possibility that the recCA already was showed some engineering,

This refers to a previous analysis Dr. Weissman had discussed about RecCA informed by data available through 2022. It is not up to date with data and analysis discussed in my manuscript. It is unclear what "recCA already was showed some engineering" is meant to imply.

In response to a comment that, "They refuted one of the weaker arguments for a laboratory leak. But didn't prove much."

> Not really. If you generously assume that the recCA has not already been influenced by engineering and that the probability of getting in his top-3 list is 100%, you get P(recCA has pattern)=4/81. But after ~400 syn muts to SC2, P(pattern)
>
> -> ~2.2% < P(pattern|DEFUSE drafts)

See above.

### “Libertarian_Virologist” (@ban_epp_gofroc)

An anonymous user falsely accused me twice of research misconduct:

> Hensel focusing on 2 pangolin CoVs with uniform spacing shows he doesn’t get the point. This pattern can occur by chance—finding a couple cases where it does is the equivalent of saying a 5% false discovery rate means hypothesis testing is a bogus method of statistical inference.

These are the only two cases I considered. These are the only two SARS-CoV-2-like, non-bat viruses with high-coverage, full-length genomes available.

> Shorter Hensel: “I Worobey around with how many restriction enzymes you ought to consider until I obtain my desired result because clearly I can’t prove this pattern is common with BsmBI and BsaI.”

I used the same set of restriction enzymes pre-specified by Bruttel *et al.* in their R script here:

```
Type2Res <- c('BbsI','BfuAI','BspQI','PaqCI','SapI','BsaI','BsmBI','BglI')
```

I did not spend time on the BsaI/BsmBI-specific analysis in Bruttel *et al.* because the analysis was inappropriate. Bruttel *et al.* wrote said that “to avoid losing power with multiple comparisons, we focus our analysis on the BsaI/BsmBI sites,” analyzing a pattern that first-author Valentin Bruttel had already selected as being the most anomalous prior to this analysis, out of a large number of equivalent restriction maps that could have been tested.

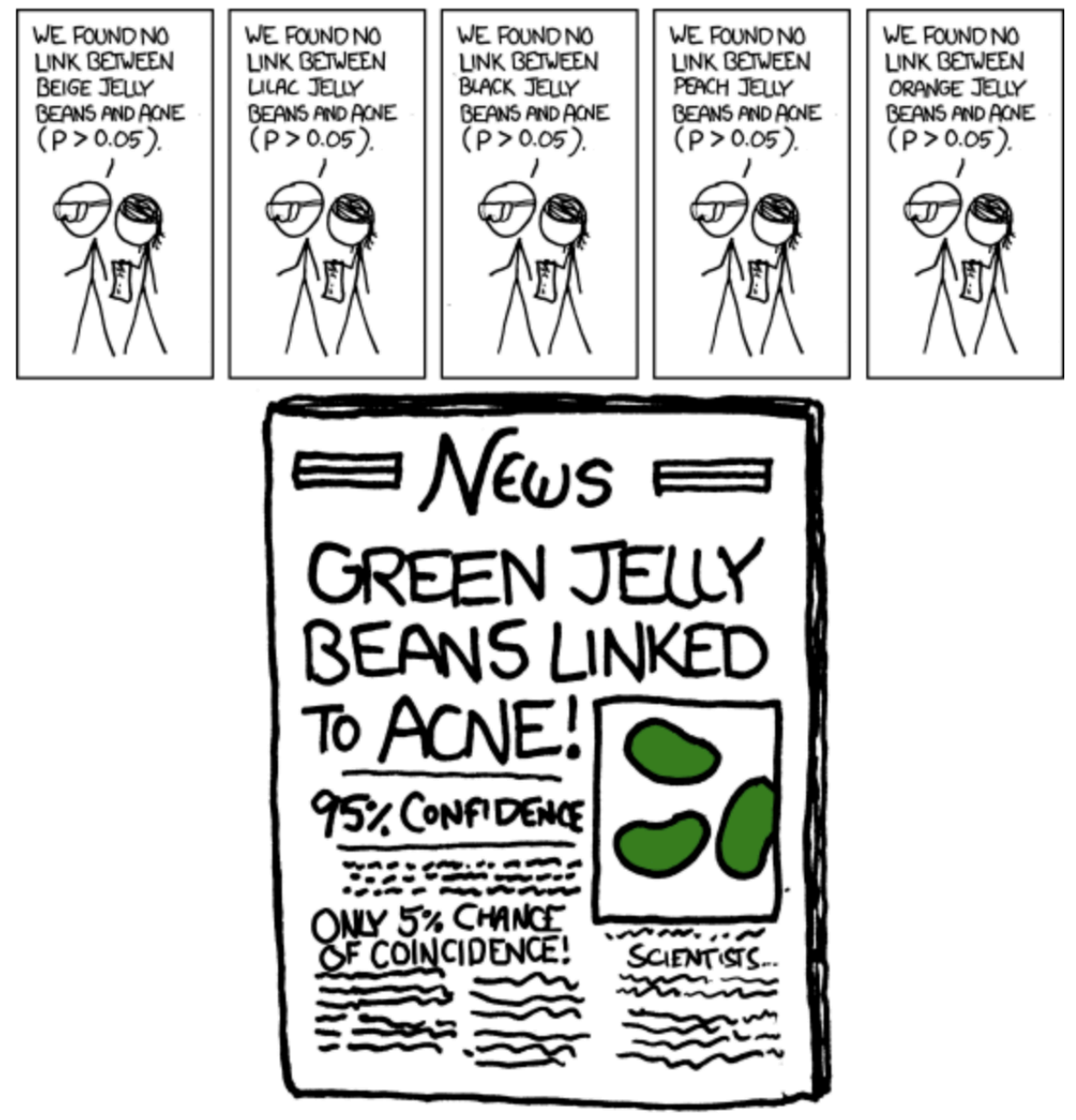

This is an inappropriate way to gain “statistical power” as demonstrated here in the seminal work, “Significant” by xkcd, available at [https://xkcd.com/882/](https://xkcd.com/882/) and excerpted here under the terms of the [Creative Commons Attribution-NonCommercial 2.5 License](https://creativecommons.org/licenses/by-nc/2.5/).

It is worth noting that a proper analysis of this type should consider a more complete list of type IIS restriction enzymes and also consider assembly methods in which internal sites are irrelevant in assembly (e.g., a BsaI site in the middle of a fragment digested by BsmBI would not be edited). However, since I showed that the observation by Bruttel *et al.* was, in fact, not remotely anomalous even by their own criteria, it was not necessary to consider the other equivalent hypotheses that they also neglected.

**On the distributions of restriction sites in human and pangolin sarbecoviruses**

Z. Hensel

## Data Supplement 1

The following is the analysis output corresponding to the summarized data described in the text and in **Table 1**. Only 20 nucleotides before and after the site of interest are shown from the alignments, but at least 100 nucleotides (at least 50 before and after each site) were analyzed. Nucleotides at contested sites are colored blue if they encode a BsaI or BsmBI restriction site and colored red otherwise.

**Position 2193-2198:** SARSCoV2 (present), RmBANAL247 (absent), RmBANAL52 (present), RaTG13 (absent)

```
Top match #1: RpBANAL103 (98.2% identity in 166 nucleotides around site)
SARSCoV2        : GGAAGGTGTAGAGTTTCTTAGAGACGGTTGGGAAATTGTTAAATTT
                  **********************************************
RpBANAL103      : GGAAGGTGTAGAGTTTCTTAGAGACGGTTGGGAAATTGTTAAATTT

Top match #2: RpYN06 (97.6% identity in 166 nucleotides around site)
SARSCoV2        : GGAAGGTGTAGAGTTTCTTAGAGACGGTTGGGAAATTGTTAAATTT
                  **********************************************
RpYN06          : GGAAGGTGTAGAGTTTCTTAGAGACGGTTGGGAAATTGTTAAATTT

Top match #3: RacCS203 (96.4% identity in 166 nucleotides around site)
SARSCoV2        : GGAAGGTGTAGAGTTTCTTAGAGACGGTTGGGAAATTGTTAAATTT
                  ************************************* ********
RacCS203        : GGAAGGTGTAGAGTTTCTTAGAGACGGTTGGGAAATTATTAAATTT
```

**Position 9751-9756:** SARSCoV2 (present), RmBANAL247 (present), RmBANAL52 (present), RaTG13 (absent)

```
Top match #1: RpYN06 (99.4% identity in 180 nucleotides around site)
SARSCoV2        : TTCTTTAGTAATTACCTAAAGAGACGTGTAGTCTTTAATGGTGTTT
                  **********************************************
RpYN06          : TTCTTTAGTAATTACCTAAAGAGACGTGTAGTCTTTAATGGTGTTT

Top match #2: RmYN02 (96.7% identity in 180 nucleotides around site)
SARSCoV2        : TTCTTTAGTAATTACCTAAAGAGACGTGTAGTCTTTAATGGTGTTT
                  ***** ** ** **********************************
RmYN02          : TTCTTCAGCAACTACCTAAAGAGACGTGTAGTCTTTAATGGTGTTT

Top match #3: RpPrC31 (96.7% identity in 180 nucleotides around site)
SARSCoV2        : TTCTTTAGTAATTACCTAAAGAGACGTGTAGTCTTTAATGGTGTTT
                  ******** ** **********************************
RpPrC31         : TTCTTTAGCAACTACCTAAAGAGACGTGTAGTCTTTAATGGTGTTT
```

**Position 10444-10449:** SARSCoV2 (absent), RmBANAL247 (present), RmBANAL52 (present), RaTG13 (present)

```
Top match #1: RpYN06 (98.5% identity in 264 nucleotides around site)
SARSCoV2        : GGTGTTTACCAATGTGCTATGAGGCCCAATTTCACTATTAAGGGTT
                  ************************** *******************
RpYN06          : GGTGTTTACCAATGTGCTATGAGGCCTAATTTCACTATTAAGGGTT

Top match #2: RpBANAL103 (96.6% identity in 264 nucleotides around site)
SARSCoV2        : GGTGTTTACCAATGTGCTATGAGGCCCAATTTCACTATTAAGGGTT
                  ******** ************** ** ** ** ** **********
RpBANAL103      : GGTGTTTATCAATGTGCTATGAGACCTAACTTTACAATTAAGGGTT

Top match #3: RmYN02 (96.6% identity in 264 nucleotides around site)
SARSCoV2        : GGTGTTTACCAATGTGCTATGAGGCCCAATTTCACTATTAAGGGTT
                  ******** ************** ** ** ** ** **********
RmYN02          : GGTGTTTATCAATGTGCTATGAGACCTAACTTTACAATTAAGGGTT
```

**Position 11648-11653:** SARSCoV2 (absent), RmBANAL247 (present), RmBANAL52 (present), RaTG13 (present)

```
Top match #1: RaTG13 (98.0% identity in 100 nucleotides around site)
SARSCoV2        : ATTTTTGTACTTGTTACTTTGGCCTCTTTTGTTTACTCAACCGCTA
                  **** ***************** ******************** **
RaTG13          : ATTTCTGTACTTGTTACTTTGGTCTCTTTTGTTTACTCAACCGTTA

Top match #2: RacCS203 (98.0% identity in 100 nucleotides around site)
SARSCoV2        : ATTTTTGTACTTGTTACTTTGGCCTCTTTTGTTTACTCAACCGCTA
                  **** ***************** ** ********************
RacCS203        : ATTTCTGTACTTGTTACTTTGGTCTTTTTTGTTTACTCAACCGCTA

Top match #3: RpPrC31 (98.0% identity in 100 nucleotides around site)
SARSCoV2        : ATTTTTGTACTTGTTACTTTGGCCTCTTTTGTTTACTCAACCGCTA
                  **** ************************************** **
RpPrC31         : ATTTCTGTACTTGTTACTTTGGCCTCTTTTGTTTACTCAACCGTTA
```

**Position 17329-17334:** SARSCoV2 (present), RmBANAL247 (present), RmBANAL52 (absent), RaTG13 (present)

```
Top match #1: Ra22QT77 (99.0% identity in 100 nucleotides around site)
```

```
SARSCoV2        : GTACTGTAAATGCATTGCCTGAGACGACAGCAGATATAGTTGTCTT
                  **************************** *****************
Ra22QT77        : GTACTGTAAATGCATTGCCTGAGACGACTGCAGATATAGTTGTCTT

Top match #2: RshSTT200 (99.0% identity in 100 nucleotides around site)
SARSCoV2        : GTACTGTAAATGCATTGCCTGAGACGACAGCAGATATAGTTGTCTT
                  **************************** *****************
RshSTT200       : GTACTGTAAATGCATTGCCTGAGACGACTGCAGATATAGTTGTCTT

Top match #3: RmBANAL247 (98.0% identity in 100 nucleotides around site)
SARSCoV2        : GTACTGTAAATGCATTGCCTGAGACGACAGCAGATATAGTTGTCTT
                  **************************** *****************
RmBANAL247      : GTACTGTAAATGCATTGCCTGAGACGACTGCAGATATAGTTGTCTT

Position 17972-17977: SARSCoV2 (present), RmBANAL247 (present), RmBANAL52 (absent), RaTG13 (present)
Top match #1: RpYN06 (99.5% identity in 220 nucleotides around site)
SARSCoV2        : ACTTTGCATAATGTCTGATAGAGACCTTTATGACAAGTTGCAATTT
                  **********************************************
RpYN06          : ACTTTGCATAATGTCTGATAGAGACCTTTATGACAAGTTGCAATTT

Top match #2: RmBANAL247 (99.1% identity in 220 nucleotides around site)
SARSCoV2        : ACTTTGCATAATGTCTGATAGAGACCTTTATGACAAGTTGCAATTT
                  **********************************************
RmBANAL247      : ACTTTGCATAATGTCTGATAGAGACCTTTATGACAAGTTGCAATTT

Top match #3: RmYN02 (98.2% identity in 220 nucleotides around site)
SARSCoV2        : ACTTTGCATAATGTCTGATAGAGACCTTTATGACAAGTTGCAATTT
                  ****** ***************** *********************
RmYN02          : ACTTTGTATAATGTCTGATAGAGATCTTTATGACAAGTTGCAATTT

Position 22922-22927: SARSCoV2 (absent), RmBANAL247 (absent), RmBANAL52 (absent), RaTG13 (present)
Top match #1: RmBANAL52 (98.0% identity in 100 nucleotides around site)
SARSCoV2        : GTAATTATAATTACCTGTATAGATTGTTTAGGAAGTCTAATCTCAA
                  **********************************************
RmBANAL52       : GTAATTATAATTACCTGTATAGATTGTTTAGGAAGTCTAATCTCAA

Top match #2: RmaBANAL236 (98.0% identity in 100 nucleotides around site)
SARSCoV2        : GTAATTATAATTACCTGTATAGATTGTTTAGGAAGTCTAATCTCAA
                  **********************************************
RmaBANAL236     : GTAATTATAATTACCTGTATAGATTGTTTAGGAAGTCTAATCTCAA

Top match #3: RpBANAL103 (98.0% identity in 100 nucleotides around site)
SARSCoV2        : GTAATTATAATTACCTGTATAGATTGTTTAGGAAGTCTAATCTCAA
                  **********************************************
RpBANAL103      : GTAATTATAATTACCTGTATAGATTGTTTAGGAAGTCTAATCTCAA

Position 22923-22928: SARSCoV2 (absent), RmBANAL247 (present), RmBANAL52 (absent), RaTG13 (absent)
Top match #1: RmBANAL52 (98.0% identity in 100 nucleotides around site)
SARSCoV2        : TAATTATAATTACCTGTATAGATTGTTTAGGAAGTCTAATCTCAAA
                  **********************************************
RmBANAL52       : TAATTATAATTACCTGTATAGATTGTTTAGGAAGTCTAATCTCAAA

Top match #2: RmaBANAL236 (98.0% identity in 100 nucleotides around site)
SARSCoV2        : TAATTATAATTACCTGTATAGATTGTTTAGGAAGTCTAATCTCAAA
                  **********************************************
RmaBANAL236     : TAATTATAATTACCTGTATAGATTGTTTAGGAAGTCTAATCTCAAA

Top match #3: RpBANAL103 (98.0% identity in 100 nucleotides around site)
SARSCoV2        : TAATTATAATTACCTGTATAGATTGTTTAGGAAGTCTAATCTCAAA
                  **********************************************
RpBANAL103      : TAATTATAATTACCTGTATAGATTGTTTAGGAAGTCTAATCTCAAA

Position 23292-23297: SARSCoV2 (absent), RmBANAL247 (present), RmBANAL52 (absent), RaTG13 (absent)
Top match #1: RmBANAL52 (96.8% identity in 124 nucleotides around site)
SARSCoV2        : TGACACTACTGATGCTGTCCGTGATCCACAGACACTTGAGATTCTT
                   *********************************************
RmBANAL52       : AGACACTACTGATGCTGTCCGTGATCCACAGACACTTGAGATTCTT

Top match #2: RmaBANAL236 (96.8% identity in 124 nucleotides around site)
SARSCoV2        : TGACACTACTGATGCTGTCCGTGATCCACAGACACTTGAGATTCTT
                   *********************************************
RmaBANAL236     : AGACACTACTGATGCTGTCCGTGATCCACAGACACTTGAGATTCTT

Top match #3: RpBANAL103 (96.8% identity in 124 nucleotides around site)
SARSCoV2        : TGACACTACTGATGCTGTCCGTGATCCACAGACACTTGAGATTCTT
                   *********************************************
RpBANAL103      : AGACACTACTGATGCTGTCCGTGATCCACAGACACTTGAGATTCTT
```

**Position 24102-24107:** SARSCoV2 (present), RmBANAL247 (present), RmBANAL52 (absent), RaTG13 (absent)

```
Top match #1: Rp22DB159 (98.0% identity in 100 nucleotides around site)
SARSCoV2       : CCTTGGTGATATTGCTGCTAGAGACCTCATTTGTGCACAAAAGTTT
                 **********************************************
Rp22DB159      : CCTTGGTGATATTGCTGCTAGAGACCTCATTTGTGCACAAAAGTTT

Top match #2: RmBtSY2 (98.0% identity in 100 nucleotides around site)
SARSCoV2       : CCTTGGTGATATTGCTGCTAGAGACCTCATTTGTGCACAAAAGTTT
                 **********************************************
RmBtSY2        : CCTTGGTGATATTGCTGCTAGAGACCTCATTTGTGCACAAAAGTTT

Top match #3: RshSTT200 (97.0% identity in 100 nucleotides around site)
SARSCoV2       : CCTTGGTGATATTGCTGCTAGAGACCTCATTTGTGCACAAAAGTTT
                 ************************ *********************
RshSTT200      : CCTTGGTGATATTGCTGCTAGAGATCTCATTTGTGCACAAAAGTTT
```

**Position 24509-24514:** SARSCoV2 (absent), RmBANAL247 (absent), RmBANAL52 (absent), RaTG13 (present)

```
Top match #1: RmBANAL52 (95.8% identity in 192 nucleotides around site)
SARSCoV2       : TTTTAAATGATATCCTTTCACGTCTTGACAAAGTTGAGGCTGAAGT
                 * ******** ***********************************
RmBANAL52      : TGTTAAATGACATCCTTTCACGTCTTGACAAAGTTGAGGCTGAAGT

Top match #2: RmaBANAL236 (95.8% identity in 192 nucleotides around site)
SARSCoV2       : TTTTAAATGATATCCTTTCACGTCTTGACAAAGTTGAGGCTGAAGT
                 * ******** ***********************************
RmaBANAL236    : TGTTAAATGACATCCTTTCACGTCTTGACAAAGTTGAGGCTGAAGT

Top match #3: RpBANAL103 (95.8% identity in 192 nucleotides around site)
SARSCoV2       : TTTTAAATGATATCCTTTCACGTCTTGACAAAGTTGAGGCTGAAGT
                 * ******** ***********************************
RpBANAL103     : TGTTAAATGACATCCTTTCACGTCTTGACAAAGTTGAGGCTGAAGT
```

## Data Supplement 2

The following is the analysis output corresponding to the summarized data described in the text and the final column of **Table 1**.

**Position:** Sequences with haplotypes matching SARS-CoV-2 described in Ou et al. 2025

**2193-2198:** RmaSTT500, RpuSTT361

**9751-9756:** RshSTT287, RshSTT219, RshSTT241, RmaSTT500, RshSTT419, RshSTT227, RshSTT268, RshSTT245, RmiSTT249, RshSTT564, RshSTT515, RshSTT610, RshSTT322, RshSTT344, RmiSTT521, RpuSTT361, RshSTT570, RshSTT336, RshSTT381, RshSTT039, RshSTT334, RshSTT317, RshSTT395, RmiSTT243, RshSTT671, RshSTT779, RshSTT494, RshSTT314

**10444-10449:** None

**11648-11653:** None

**17329-17334:** RshSTT287, RshSTT219, RacSTT347, RacSTT333, RshSTT241, RmaSTT500, RshSTT419, RshSTT227, RshSTT268, RshSTT245, RmiSTT249, RshSTT564, RshSTT515, RshSTT610, RshSTT322, RshSTT344, RmiSTT521, RpuSTT361, RshSTT570, RshSTT336, RshSTT381, RshSTT039, RacSTT334, RshSTT317, RshSTT395, RmiSTT243, RacSTT351, RshSTT671, RshSTT779, RshSTT494, RacSTT345, RshSTT314

**17972-17977:** None

**22922-22927:** None

**22923-22928:** None

**23292-23297:** RacSTT334, RacSTT347, RacSTT333, RacSTT345, RacSTT351, RpuSTT361

**24102-24107:** RmaSTT500

**24509-24514:** RacSTT334, RacSTT347, RacSTT333, RacSTT345, RmaSTT500, RacSTT351, RpuSTT361

## Data Supplement 3

The following is the analysis output corresponding to the analysis of BsaI recognition sites for pangolin coronavirus MjGuangdong. Only 20 nucleotides before and after the site of interest are shown from the alignments, but at least 100 nucleotides (at least 50 before and after each site) were analyzed. Nucleotides at contested sites are colored blue if they encode a BsaI recognition site and colored red otherwise.

```
Position 7558-7563: MjGuangdong (present), RmBANAL247 (absent), RmBANAL52 (absent), RaTG13 (absent)
Top match #1: Rp22DB159 (88.0% identity in 234 nucleotides around site)
MjGuangdong    : GACTTATCATTACAGTTTAAGAGACCAATTAACCCAACTGACCAGT
                 *** ***** ********** ******** ** ** ** *******
Rp22DB159      : GACCTATCACTACAGTTTAAAAGACCAATAAATCCTACCGACCAGT

Top match #2: RmBtSY2 (88.0% identity in 234 nucleotides around site)
MjGuangdong    : GACTTATCATTACAGTTTAAGAGACCAATTAACCCAACTGACCAGT
                 *** ***** ********** ******** ** ** ** *******
RmBtSY2        : GACCTATCACTACAGTTTAAAAGACCAATAAATCCTACCGACCAGT

Top match #3: RpYN2021 (88.0% identity in 234 nucleotides around site)
MjGuangdong    : GACTTATCATTACAGTTTAAGAGACCAATTAACCCAACTGACCAGT
                 *** ***** ********** ******** ** ** ** *******
RpYN2021       : GACCTATCACTACAGTTTAAAAGACCAATAAATCCTACCGACCAGT

Position 10303-10308: MjGuangdong (present), RmBANAL247 (present), RmBANAL52 (present), RaTG13 (present)
Top match #1: MjGuangxi (94.0% identity in 100 nucleotides around site)
MjGuangdong    : GGTGTTTACCAATGTGCTATGAGACCTAATTTTACCATCAAAGGTT
                 *********** *********************** ** ***** *
MjGuangxi      : GGTGTTTACCAGTGTGCTATGAGACCTAATTTTACTATTAAAGGAT

Top match #2: RpYN06 (92.0% identity in 100 nucleotides around site)
MjGuangdong    : GGTGTTTACCAATGTGCTATGAGACCTAATTTTACCATCAAAGGTT
                 *********************** ******** ** ** ** ****
RpYN06         : GGTGTTTACCAATGTGCTATGAGGCCTAATTTCACTATTAAGGGTT

Top match #3: RpHN2021G (92.0% identity in 100 nucleotides around site)
MjGuangdong    : GGTGTTTACCAATGTGCTATGAGACCTAATTTTACCATCAAAGGTT
                 ***************** ***************** ** ** ****
RpHN2021G      : GGTGTTTACCAATGTGCCATGAGACCTAATTTTACTATTAAGGGTT

Position 11507-11512: MjGuangdong (absent), RmBANAL247 (present), RmBANAL52 (present), RaTG13 (present)
Top match #1: RmBANAL52 (96.2% identity in 318 nucleotides around site)
MjGuangdong    : ATTTCTGTACTTGTTACTTTGGCCTCTTCTGTTTACTCAACCGCTA
                 ********************** ***** *****************
RmBANAL52      : ATTTCTGTACTTGTTACTTTGGTCTCTTTTGTTTACTCAACCGCTA

Top match #2: RmaBANAL236 (96.2% identity in 318 nucleotides around site)
MjGuangdong    : ATTTCTGTACTTGTTACTTTGGCCTCTTCTGTTTACTCAACCGCTA
                 ********************** ***** *****************
RmaBANAL236    : ATTTCTGTACTTGTTACTTTGGTCTCTTTTGTTTACTCAACCGCTA

Top match #3: RmBANAL247 (96.2% identity in 318 nucleotides around site)
MjGuangdong    : ATTTCTGTACTTGTTACTTTGGCCTCTTCTGTTTACTCAACCGCTA
                 ********************** ***** *****************
RmBANAL247     : ATTTCTGTACTTGTTACTTTGGTCTCTTTTGTTTACTCAACCGCTA

Position 16522-16527: MjGuangdong (present), RmBANAL247 (absent), RmBANAL52 (absent), RaTG13 (absent)
Top match #1: RsHN2021E (87.0% identity in 100 nucleotides around site)
MjGuangdong    : AAACATTGAAAGCAACAGAAGAGACCTTTAAACTATCTTACGGCAT
                 ****  * ***** ** ******** ** *********** *****
RsHN2021E      : AAACGCTCAAAGCTACTGAAGAGACATTCAAACTATCTTATGGCAT

Top match #2: RsHN2021F (87.0% identity in 100 nucleotides around site)
MjGuangdong    : AAACATTGAAAGCAACAGAAGAGACCTTTAAACTATCTTACGGCAT
```

```
                 ****  * ***** ** ******** ** *********** *****
RsHN2021F      : AAACGCTCAAAGCTACTGAAGAGACATTCAAACTATCTTATGGCAT

Top match #3: RbBM4831 (86.0% identity in 100 nucleotides around site)
MjGuangdong    : AAACATTGAAAGCAACAGAAGAGACCTTTAAACTATCTTACGGCAT
                 **** ** ***** *  ******** *********** ** *****
RbBM4831       : AAACTTTAAAAGCTAATGAAGAGACATTTAAACTATCCTATGGCAT
```

**Position 17831-17836:** MjGuangdong (present), RmBANAL247 (present), RmBANAL52 (absent), RaTG13 (present)

```
Top match #1: RaTG13 (96.0% identity in 100 nucleotides around site)
MjGuangdong    : ACTTTGCATAATGTCAGATAGAGACCTTTATGACAAGTTGCAATTT
                 *************** ******************************
RaTG13         : ACTTTGCATAATGTCTGATAGAGACCTTTATGACAAGTTGCAATTT

Top match #2: SARSCoV2 (96.0% identity in 100 nucleotides around site)
MjGuangdong    : ACTTTGCATAATGTCAGATAGAGACCTTTATGACAAGTTGCAATTT
                 *************** ******************************
SARSCoV2       : ACTTTGCATAATGTCTGATAGAGACCTTTATGACAAGTTGCAATTT

Top match #3: RpYN06 (96.0% identity in 100 nucleotides around site)
MjGuangdong    : ACTTTGCATAATGTCAGATAGAGACCTTTATGACAAGTTGCAATTT
                 *************** ******************************
RpYN06         : ACTTTGCATAATGTCTGATAGAGACCTTTATGACAAGTTGCAATTT
```

**Position 18326-18331:** MjGuangdong (present), RmBANAL247 (absent), RmBANAL52 (absent), RaTG13 (absent)

```
Top match #1: Ra22QT77 (92.0% identity in 100 nucleotides around site)
MjGuangdong    : TAGTGCTAAGCCACCACCTGGAGACCAGTTTAAACATCTTATACCA
                 ********* ************** *********************
Ra22QT77       : TAGTGCTAAACCACCACCTGGAGATCAGTTTAAACATCTTATACCA

Top match #2: RshSTT200 (92.0% identity in 100 nucleotides around site)
MjGuangdong    : TAGTGCTAAGCCACCACCTGGAGACCAGTTTAAACATCTTATACCA
                 ********* ************** *********************
RshSTT200      : TAGTGCTAAACCACCACCTGGAGATCAGTTTAAACATCTTATACCA

Top match #3: RacCS203 (92.0% identity in 100 nucleotides around site)
MjGuangdong    : TAGTGCTAAGCCACCACCTGGAGACCAGTTTAAACATCTTATACCA
                 ************************ *********************
RacCS203       : TAGTGCTAAGCCACCACCTGGAGATCAGTTTAAACATCTTATACCA
```

**Position 22157-22162:** MjGuangdong (present), RmBANAL247 (absent), RmBANAL52 (absent), RaTG13 (absent)

```
Top match #1: RpBANAL103 (80.4% identity in 112 nucleotides around site)
MjGuangdong    : ACTCCTCACTATACATAGAGGAGACCCCATGCCT------AATAAT
                 ****** ** ** *********************************
RpBANAL103     : ACTCCTTACAATTCATAGAGGAGACCCCATGCCT------AATAAT

Top match #2: RmaBANAL236 (79.5% identity in 112 nucleotides around site)
MjGuangdong    : ACTCCTCACTATACATAGAGGAGACCCCATGCCT------AATAAT
                 *** ***** ** *********************************
RmaBANAL236    : ACTTCTCACAATTCATAGAGGAGACCCCATGCCT------AATAAT

Top match #3: RpVZXC21 (79.5% identity in 112 nucleotides around site)
MjGuangdong    : ACTCCTCACTATACATAGAGGAGACCCCATGCCT------AATAAT
                 ************ ************** *** **************
RpVZXC21       : ACTCCTCACTATTCATAGAGGAGACCCTATGTCT------AATAAT
```

**Position 22769-22774:** MjGuangdong (absent), RmBANAL247 (present), RmBANAL52 (absent), RaTG13 (absent)

```
Top match #1: RmBANAL52 (90.0% identity in 100 nucleotides around site)
MjGuangdong    : TAATTATAACTACCTTTATAGATTGTTTAGAAAGTCCAACCTCAAA
                 ********* ***** ************** ***** ** ******
RmBANAL52      : TAATTATAATTACCTGTATAGATTGTTTAGGAAGTCTAATCTCAAA

Top match #2: RmaBANAL236 (90.0% identity in 100 nucleotides around site)
MjGuangdong    : TAATTATAACTACCTTTATAGATTGTTTAGAAAGTCCAACCTCAAA
                 ********* ***** ************** ***** ** ******
RmaBANAL236    : TAATTATAATTACCTGTATAGATTGTTTAGGAAGTCTAATCTCAAA
```

```
Top match #3: RpBANAL103 (90.0% identity in 100 nucleotides around site)
MjGuangdong     : TAATTATAACTACCTTTATAGATTGTTTAGAAAGTCCAACCTCAAA
                  ********* ***** ************** ***** ** ******
RpBANAL103      : TAATTATAATTACCTGTATAGATTGTTTAGGAAGTCTAATCTCAAA

Position 23138-23143: MjGuangdong (absent), RmBANAL247 (present), RmBANAL52 (absent), RaTG13 (absent)
Top match #1: SARSCoV2 (92.7% identity in 288 nucleotides around site)
MjGuangdong     : CGACACTACTGATGCTGTCCGTGATCCACAGACACTTGAAATTCTT
                   ************************************** ******
SARSCoV2        : TGACACTACTGATGCTGTCCGTGATCCACAGACACTTGAGATTCTT

Top match #2: RmaBANAL236 (89.9% identity in 288 nucleotides around site)
MjGuangdong     : CGACACTACTGATGCTGTCCGTGATCCACAGACACTTGAAATTCTT
                   ************************************** ******
RmaBANAL236     : AGACACTACTGATGCTGTCCGTGATCCACAGACACTTGAGATTCTT

Top match #3: RpBANAL103 (89.9% identity in 288 nucleotides around site)
MjGuangdong     : CGACACTACTGATGCTGTCCGTGATCCACAGACACTTGAAATTCTT
                   ************************************** ******
RpBANAL103      : AGACACTACTGATGCTGTCCGTGATCCACAGACACTTGAGATTCTT

Position 23936-23941: MjGuangdong (absent), RmBANAL247 (present), RmBANAL52 (absent), RaTG13 (absent)
Top match #1: RmBANAL52 (98.0% identity in 100 nucleotides around site)
MjGuangdong     : CCTTGGTGATATTGCCGCTAGAGATCTTATTTGTGCACAAAAGTTT
                  *************** ******************** *********
RmBANAL52       : CCTTGGTGATATTGCTGCTAGAGATCTTATTTGTGCTCAAAAGTTT

Top match #2: RmaBANAL236 (98.0% identity in 100 nucleotides around site)
MjGuangdong     : CCTTGGTGATATTGCCGCTAGAGATCTTATTTGTGCACAAAAGTTT
                  *************** ******************** *********
RmaBANAL236     : CCTTGGTGATATTGCTGCTAGAGATCTTATTTGTGCTCAAAAGTTT

Top match #3: RpBANAL103 (98.0% identity in 100 nucleotides around site)
MjGuangdong     : CCTTGGTGATATTGCCGCTAGAGATCTTATTTGTGCACAAAAGTTT
                  *************** ******************** *********
RpBANAL103      : CCTTGGTGATATTGCTGCTAGAGATCTTATTTGTGCTCAAAAGTTT
```

## Data Supplement 4

The following is the analysis output corresponding to the analysis of PaqCI recognition sites for pangolin coronavirus MjGuangxi. Only 20 nucleotides before and after the site of interest are shown from the alignments, but at least 100 nucleotides (at least 50 before and after each site) were analyzed. Nucleotides at contested sites are colored blue if they encode a PaqCI recognition site and colored red otherwise.

```
Position 4137-4143: MjGuangxi (present), RmBANAL247 (absent), RmBANAL52 (absent), RaTG13 (absent)
Top match #1: RmYN07 (73.0% identity in 100 nt window)
MjGuangxi      : TTGTCATACCAACAAAGAAAGCAGGTGGTACTACAGAAATGCTTGCAAAGGCA
                 **** *****  * ***** ** *************************  ***
RmYN07         : TTGTAATACCTTCTAAGAAGGCTGGTGGTACTACAGAAATGCTTGCAAGAGCA

Top match #2: RpF46 (72.0% identity in 100 nt window)
MjGuangxi      : TTGTCATACCAACAAAGAAAGCAGGTGGTACTACAGAAATGCTTGCAAAGGCA
                 **** *****  * ***** ** *********** *************  ***
RpF46          : TTGTAATACCTTCTAAGAAGGCTGGTGGTACTACGGAAATGCTTGCAAGAGCA

Top match #3: RaLYRa11 (72.0% identity in 100 nt window)
MjGuangxi      : TTGTCATACCAACAAAGAAAGCAGGTGGTACTACAGAAATGCTTGCAAAGGCA
                 **** *****  * ***** ** *********** *************  ***
RaLYRa11       : TTGTAATACCTTCTAAGAAGGCTGGTGGTACTACGGAAATGCTTGCAAGAGCA

Position 5299-5305: MjGuangxi (absent), RmBANAL247 (present), RmBANAL52 (present), RaTG13 (present)
Top match #1: MjGuangdong (82.0% identity in 100 nt window)
MjGuangxi      : GATAGAATTGAAGTTTAATCCACCAGCATTGCAAGACGCCTACTACAGGGCTA
                  ***** ***************** ** ******** ** ** ** ** ****
MjGuangdong    : AATAGAGTTGAAGTTTAATCCACCTGCTTTGCAAGATGCTTATTATAGAGCTA

Top match #2: RaTG13 (81.0% identity in 100 nt window)
MjGuangxi      : GATAGAATTGAAGTTTAATCCACCAGCATTGCAAGACGCCTACTACAGGGCTA
                  ***** ***** ***** ***** ** ** ***** ************** *
RaTG13         : AATAGAGTTGAAATTTAACCCACCTGCTTTACAAGATGCCTACTACAGGGCAA

Top match #3: RpVZXC21 (81.0% identity in 100 nt window)
MjGuangxi      : GATAGAATTGAAGTTTAATCCACCAGCATTGCAAGACGCCTACTACAGGGCTA
                  ***** ***** ************** ** ***** ***** ** ***** *
RpVZXC21       : AATAGAGTTGAAATTTAATCCACCAGCTTTACAAGATGCCTATTATAGGGCAA

Position 5653-5659: MjGuangxi (present), RmBANAL247 (present), RmBANAL52 (present), RaTG13 (present)
Top match #1: RacCS203 (90.0% identity in 110 nt window)
MjGuangxi      : TTTTGTTATGATGTCTGCACCACCTGCTGAATATAAACTAAAGCATGGTACTT
                 ************************** ******* ** ***************
RacCS203       : TTTTGTTATGATGTCTGCACCACCTGTTGAATATGAATTAAAGCATGGTACTT

Top match #2: RpHN2021A (90.0% identity in 110 nt window)
MjGuangxi      : TTTTGTTATGATGTCTGCACCACCTGCTGAATATAAACTAAAGCATGGTACTT
                 **************************** ***** **** ********* * *
RpHN2021A      : TTTTGTTATGATGTCTGCACCACCTGCTCAATATGAACTTAAGCATGGTGCAT

Top match #3: RpVZXC21 (90.0% identity in 110 nt window)
MjGuangxi      : TTTTGTTATGATGTCTGCACCACCTGCTGAATATAAACTAAAGCATGGTACTT
                 ************************ **  ***** **** *********** *
RpVZXC21       : TTTTGTTATGATGTCTGCACCACCCGCCCAATATGAACTTAAGCATGGTACAT

Position 9369-9375: MjGuangxi (present), RmBANAL247 (absent), RmBANAL52 (absent), RaTG13 (absent)
Top match #1: RacCS203 (85.0% identity in 100 nt window)
MjGuangxi      : ACATATCAGCTTCAATTGTTGCAGGTGGTTTAGTTGCTATATTTGTAACTTGT
                 ******* ** ** ***** ***** *** * ********  * ***** ***
RacCS203       : ACATATCTGCATCTATTGTGGCAGGAGGTATTGTTGCTATTATAGTAACGTGT

Top match #2: RpJCC9 (84.0% identity in 100 nt window)
MjGuangxi      : ACATATCAGCTTCAATTGTTGCAGGTGGTTTAGTTGCTATATTTGTAACTTGT
                 ********** ** ** ** ** ** ***** ********  * ***** **
```

```
RpJCC9         : ACATATCAGCATCTATAGTAGCTGGCGGTTTTGTTGCTATCGTAGTAACATGC

Top match #3: RpPrC31 (83.0% identity in 100 nt window)
MjGuangxi      : ACATATCAGCTTCAATTGTTGCAGGTGGTTTAGTTGCTATATTTGTAACTTGT
                 ********** ** ** ** ** ****** * ********* * ***** **
RpPrC31        : ACATATCAGCATCTATAGTAGCTGGTGGTATTGTTGCTATAGTAGTAACGTGC

Position 15071-15077: MjGuangxi (present), RmBANAL247 (absent), RmBANAL52 (absent), RaTG13 (absent)
Top match #1: Rt22SL115 (91.1% identity in 158 nt window)
MjGuangxi      : AAAATAGAGCTCGCACCGTTGCAGGTGTTTCTATTTGTAGTACTATGACTAAT
                 * ***************** ** ***** ***** ************** ***
Rt22SL115      : AGAATAGAGCTCGCACCGTAGCTGGTGTCTCTATCTGTAGTACTATGACAAAT

Top match #2: Rt22SL9 (91.1% identity in 158 nt window)
MjGuangxi      : AAAATAGAGCTCGCACCGTTGCAGGTGTTTCTATTTGTAGTACTATGACTAAT
                 * ***************** ** ***** ***** ************** ***
Rt22SL9        : AGAATAGAGCTCGCACCGTAGCTGGTGTCTCTATCTGTAGTACTATGACAAAT

Top match #3: RstYN03 (91.1% identity in 158 nt window)
MjGuangxi      : AAAATAGAGCTCGCACCGTTGCAGGTGTTTCTATTTGTAGTACTATGACTAAT
                 * ******************** ***** ***** ************** ***
RstYN03        : AGAATAGAGCTCGCACCGTTGCTGGTGTCTCTATCTGTAGTACTATGACCAAT

Position 16567-16573: MjGuangxi (present), RmBANAL247 (absent), RmBANAL52 (absent), RaTG13 (absent)
Top match #1: RhGB01 (89.2% identity in 102 nt window)
MjGuangxi      : GGTGATTACATTCTTGCGAACACCTGCACAGAAAGACTTAAACTTTTTGCTGC
                 *********** ***** ******** ******* ********* ********
RhGB01         : GGTGATTACATACTTGCTAACACCTGTACAGAAAAACTTAAACTCTTTGCTGC

Top match #2: MjGuangdong (89.2% identity in 102 nt window)
MjGuangxi      : GGTGATTACATTCTTGCGAACACCTGCACAGAAAGACTTAAACTTTTTGCTGC
                 ***************** ******** ** ** *********** ** *****
MjGuangdong    : GGTGATTACATTCTTGCTAACACCTGTACTGAGAGACTTAAACTGTTCGCTGC

Top match #3: Rco319 (89.2% identity in 102 nt window)
MjGuangxi      : GGTGATTACATTCTTGCGAACACCTGCACAGAAAGACTTAAACTTTTTGCTGC
                 ***************** *********** *********** ******** **
Rco319         : GGTGATTACATTCTTGCTAACACCTGCACTGAAAGACTTAAGCTTTTTGCAGC

Position 23410-23416: MjGuangxi (present), RmBANAL247 (absent), RmBANAL52 (absent), RaTG13 (absent)
Top match #1: RaLYRa11 (86.5% identity in 104 nt window)
MjGuangxi      : CATTCATGCAGAACAACTTACACCTGCCTGGCGTGTTTACTCT------GCAG
                  *********** ***** ** ***** ****** ******************
RaLYRa11       : TATTCATGCAGACCAACTCACCCCTGCTTGGCGTATTTACTCT------GCAG

Top match #2: Rp22DB159 (84.6% identity in 104 nt window)
MjGuangxi      : CATTCATGCAGAACAACTTACACCTGCCTGGCGTGTTTACTCT------GCAG
                  ** ******** ******** **  * ********************* ***
Rp22DB159      : TATCCATGCAGACCAACTTACTCCCACTTGGCGTGTTTACTCT------ACAG

Top match #3: RmBtSY2 (84.6% identity in 104 nt window)
MjGuangxi      : CATTCATGCAGAACAACTTACACCTGCCTGGCGTGTTTACTCT------GCAG
                  ** ******** ******** **  * ********************* ***
RmBtSY2        : TATCCATGCAGACCAACTTACTCCCACTTGGCGTGTTTACTCT------ACAG

Position 23548-23554: MjGuangxi (absent), RmBANAL247 (present), RmBANAL52 (absent), RaTG13 (absent)
Top match #1: Rt22QB8 (80.0% identity in 100 nt window)
MjGuangxi      : CATATGTGCAAGTTACCATTCCATGTCA---TCATTG---CGTAGTGTC---A
                 *** ***** ** ****** ***  ********  * *********** ****
Rt22QB8        : CATTTGTGCTAGCTACCATACCACTTCA---TCTCTT---CGTAGTGTA---A

Top match #2: Rt22QB78 (79.0% identity in 100 nt window)
MjGuangxi      : CATATGTGCAAGTTACCATTCCATGTCA---TCATTG---CGTAGTGTC---A
                 *** ***** ** *****  ***  ********  * *********** ****
Rt22QB78       : CATTTGTGCTAGCTACCACACCACTTCA---TCTCTT---CGTAGTGTA---A

Top match #3: Rt22QT124 (79.0% identity in 100 nt window)
```

```
MjGuangxi      : CATATGTGCAAGTTACCATTCCATGTCA---TCATTG---CGTAGTGTC---A
                 *** ***** ** ****** * *  ********  * *********** ***
Rt22QT124      : CATTTGTGCTAGCTACCATACTACTTCA---TCTCTT---CGTAGTGTA---G
```

**Position 24177-24183:** MjGuangxi (absent), RmBANAL247 (absent), RmBANAL52 (present), RaTG13 (present)
Top match #1: MjGuangdong (90.5% identity in 116 nt window)

```
MjGuangxi      : CCTCAGGTTGGACCTTTGGTGCAGGAGCTGCTTTACAAATACCCTTTGCAATG
                 * ******************** ** ** ** ***** ***** ***** ***
MjGuangdong    : CATCAGGTTGGACCTTTGGTGCTGGTGCAGCATTACAGATACCATTTGCTATG
```

Top match #2: RshSTT200 (90.5% identity in 116 nt window)

```
MjGuangxi      : CCTCAGGTTGGACCTTTGGTGCAGGAGCTGCTTTACAAATACCCTTTGCAATG
                 * *********************** ***** *********** *********
RshSTT200      : CTTCAGGTTGGACCTTTGGTGCAGGTGCTGCGTTACAAATACCATTTGCAATG
```

Top match #3: RmBANAL52 (90.5% identity in 116 nt window)

```
MjGuangxi      : CCTCAGGTTGGACCTTTGGTGCAGGAGCTGCTTTACAAATACCCTTTGCAATG
                 * ** ******************** ***** *********** ***** ***
RmBANAL52      : CTTCTGGTTGGACCTTTGGTGCAGGTGCTGCATTACAAATACCATTTGCTATG
```

**Position 27029-27035:** MjGuangxi (absent), RmBANAL247 (absent), RmBANAL52 (present), RaTG13 (present)
Top match #1: RacCS203 (97.2% identity in 212 nt window)

```
MjGuangxi      : TGGGAGCTTCGCAGCGTGTAGCCGGTGACTCAGGTTTTGCTGCATACAGTCGC
                 ******* ************** ******************************
RacCS203       : TGGGAGCGTCGCAGCGTGTAGCAGGTGACTCAGGTTTTGCTGCATACAGTCGC
```

Top match #2: RmBtSY2 (96.7% identity in 212 nt window)

```
MjGuangxi      : TGGGAGCTTCGCAGCGTGTAGCCGGTGACTCAGGTTTTGCTGCATACAGTCGC
                 ********************** ***** ************************
RmBtSY2        : TGGGAGCTTCGCAGCGTGTAGCAGGTGATTCAGGTTTTGCTGCATACAGTCGC
```

Top match #3: SARSCoV2 (96.2% identity in 212 nt window)

```
MjGuangxi      : TGGGAGCTTCGCAGCGTGTAGCCGGTGACTCAGGTTTTGCTGCATACAGTCGC
                 ********************** ******************************
SARSCoV2       : TGGGAGCTTCGCAGCGTGTAGCAGGTGACTCAGGTTTTGCTGCATACAGTCGC
```